# Accelerating Large-Scale Data Exploration through Data Diffusion

Ioan Raicu[1], Yong Zhao[2], Ian Foster[1,3,4], Alex Szalay[5]

[1]Department of Computer Science, University of Chicago, IL, USA
[2]Microsoft Corporation, Redmond, WA, USA
[3]Computation Institute, University of Chicago and Argonne National Laboratory, USA
[4]Mathematics and Computer Science Division, Argonne National Laboratory, Argonne IL, USA
[5]Department of Physics and Astronomy, The Johns Hopkins University, Baltimore MD, USA

iraicu@cs.uchicago.edu, yozha@microsoft.com, foster@mcs.anl.gov, szalay@jhu.edu

## ABSTRACT

Data-intensive applications often require exploratory analysis of large datasets. If analysis is performed on distributed resources, data locality can be crucial to high throughput and performance. We propose a "data diffusion" approach that acquires compute and storage resources dynamically, replicates data in response to demand, and schedules computations close to data. As demand increases, more resources are acquired, thus allowing faster response to subsequent requests that refer to the same data; when demand drops, resources are released. This approach can provide the benefits of dedicated hardware without the associated high costs, depending on workload and resource characteristics. The approach is reminiscent of cooperative caching, web-caching, and peer-to-peer storage systems, but addresses different application demands. Other data-aware scheduling approaches assume dedicated resources, which can be expensive and/or inefficient if load varies significantly. To explore the feasibility of the data diffusion approach, we have extended the Falkon resource provisioning and task scheduling system to support data caching and data-aware scheduling. Performance results from both micro-benchmarks and a large scale astronomy application demonstrate that our approach improves performance relative to alternative approaches, as well as provides improved scalability as aggregated I/O bandwidth scales linearly with the number of data cache nodes.

## 1. INTRODUCTION AND MOTIVATION

The ability to analyze large quantities of data has become increasingly important in many fields. To achieve rapid turnaround, data may be distributed over hundreds of computers. In such circumstances, data locality has been shown to be crucial to the successful and efficient use of large distributed systems for data-intensive applications [7, 34].

One approach to achieving data locality—adopted, for example, by Google [3, 11]—is to build large compute-storage farms dedicated to storing data and responding to user requests for processing. However, such approaches can be expensive (in terms of idle resources) if load varies significantly over the two dimensions of time and/or the data of interest.

This paper proposes an alternative *data diffusion* approach, in which resources required for data analysis are acquired dynamically, in response to demand. Resources may be acquired either "locally" or "remotely"; their location only matters in terms of associated cost tradeoffs. Both data and applications are copied (they "diffuse") to newly acquired resources for processing. Acquired resources (computers and storage) and the data that they hold can be "cached" for some time, thus allowing more rapid responses to subsequent requests. If demand drops, resources can be released, allowing their use for other purposes. Thus, data diffuses over an increasing number of CPUs as demand increases, and then contracting as load reduces.

Data diffusion thus involves a combination of dynamic resource provisioning, data caching, and data-aware scheduling. The approach is reminiscent of cooperative caching [18], cooperative web-caching [19], and peer-to-peer storage systems [17]. (Other data-aware scheduling approaches tend to assume static resources [1, 2].) However, in our approach we need to acquire dynamically not only storage resources but also computing resources. In addition, datasets may be terabytes in size and data access is for analysis (not retrieval). Further complicating the situation is our limited knowledge of workloads, which may involve many different applications.

In our exploration of these issues, we build upon previous work on Falkon, a Fast and Light-weight tasK executiON framework [4, 12], which provides for dynamic acquisition and release of resources ("workers") and the dispatch of analysis tasks to those workers. We describe Falkon data caching extensions that enable (in their current instantiation) the management of tens of millions of files spanning hundreds of multiple storage resources.

In principle, data diffusion can provide the benefit of dedicated hardware without the associated high costs. It can also overcome inefficiencies that may arise when executing data-intensive applications in distributed ("grid") environments, due to the high costs of data movement [34]: if workloads have sufficient internal locality of reference [22], then it is feasible to acquire and use even remote resources despite high initial data movement costs.

The performance achieved with data diffusion depends crucially on the precise characteristics of application workloads and the underlying infrastructure. As a first step towards quantifying these dependences, we have conducted experiments with both micro-benchmarks and a large scale astronomy application. The experiments presented here do not investigate the effects of dynamic resource provisioning, which we will address in future work. They show that our approach improves performance relative to alternative approaches, and provides improved scalability as aggregated I/O bandwidth scales linearly with the number of data cache nodes.

## 2. RELATED WORK
The results presented here build on our past work on resource provisioning [12] and task dispatching [4], and implement ideas outlined in a previous short paper [26].

Data management becomes more useful if coupled with compute resource management. Ranganathan et al. used simulation studies [10] to show that proactive data replication can improve application performance. The Stork [28] scheduler seeks to improve performance and reliability when batch scheduling by explicitly scheduling data placement operations. However, while Stork can be used with other system components to co-schedule CPU and storage resources, there is no attempt to retain nodes between tasks as in our work.

The GFarm team implemented a data-aware scheduler in Gfarm using an LSF scheduler plugin [1, 23]. Their performance results are for a small system (6 nodes, 300 jobs, 900 MB input files, 2640 second workload without data-aware scheduling, 1650 seconds with data-aware scheduling, 0.1–0.2 jobs/sec, 90MB/s to 180MB/s data rates); it is not clear that it scales to larger systems. In contrast, we have tested our proposed data diffusion with 64 nodes, 100K jobs, input data ranging from 1B to 1GB, workflows exceeding 1000 jobs/sec, and data rates exceeding 8750 MB/s.

BigTable [21], Google File System (GFS) [3], and MapReduce [11] (or the open source implementation in Hadoop [27]) couple data and computing resources to accelerate data-intensive applications. However, these systems all assume a static set of resources. Furthermore, the tight coupling of execution engine (MapReduce, Hadoop) and file system (GFS) means that applications that want to use these tools must be modified. In our work, we further extend this fusion of data and compute resource management by also enabling dynamic resource provisioning, which we assert can provide performance advantages when workload characteristics change over time. In addition, because we perform data movement prior to task execution, we are able to run applications unmodified.

The batch-aware distributed file system (BAD-FS) [29] caches data transferred from centralized data storage servers to local disks. However, it uses a capacity-aware scheduler which is differentiated from a data-aware scheduler by its focus on ensuring that jobs have enough capacity to execute, rather than on placing jobs to minimize cache-to-cache transfers. We expect BAD-FS to produce more local area traffic than data diffusion. Although BAD-FS addresses dynamic deployment via multi-level scheduling, it does not address dynamic reconfiguration during the lifetime of the deployment, a key feature offered in Falkon.

## 3. DATA DIFFUSION ARCHITECTURE
We describe first the Falkon task dispatch framework [4] and then the Falkon extensions that implement data diffusion.

### 3.1 Falkon
To enable the rapid execution of many tasks on distributed resources, Falkon combines (1) multi-level scheduling [13, 14] to separate resource acquisition (via requests to batch schedulers) from task dispatch, and (2) a streamlined dispatcher to achieve one to two orders of magnitude higher throughput (487 tasks/sec) and scalability (54K executors, 2M queued tasks) than other resource managers [4]. Recent tuning and experimentation have achieved throughputs in excess of 3750 tasks/sec and the management of up to 1M simulated executors without significant degradation of throughput.

The Falkon architecture comprises a set of (dynamically allocated) *executors* that cache and analyze data; a *dynamic resource provisioner* (DRP) that manages the creation and deletion of executors; and a *dispatcher* that dispatches each incoming task to an executor. The provisioner uses tunable allocation and de-allocation policies to provision resources adaptively.

In prior work, we have assumed that each task scheduled by Falkon accessed input and output files at remote persistent storage locations, for example via a shared file system, gridFTP server, or web server. This strategy provides acceptable performance in many cases, but does not scale for data-intensive applications, such as image stacking [5, 6] and mosaic services [15] in astronomy, which access digital image datasets that are typically large (multiple terabytes) and contain many (100M+) objects stored into many (1M+) files.

### 3.2 Enhancing Falkon with Data Diffusion
The intent of data diffusion is to achieve a separation of concerns between the core logic of data-intensive applications and the complicated task of managing large data sets, while improving resource utilization and ultimately application performance. To this end, we incorporate data caches in executors and data-aware task scheduling algorithms in the dispatcher.

Individual executors manage their own caches, using local eviction policies, and communicate changes in cache content to the dispatcher. The dispatcher sends tasks to nodes that have cached the most needed data, along with the information on how to locate needed data. An executor that receives a task to execute will, if possible, access required data from its local cache or request it from peer executors. Only if no cached copy is available does the executor request a copy from persistent storage.

As in other computing systems that make use of caches, this general approach can enable significant performance improvements if an application's working set fits in faster storage.

*3.2.1 Data Diffusion Architecture*
To support location-aware scheduling, we implement a centralized index within the dispatcher that records the location of every cached data object. This index is maintained loosely coherent with the contents of the executor's caches via periodic update messages generated by the executors. In addition, each executor maintains a local index to record the location of its cached data objects. We believe that this hybrid (but essentially centralized) architecture provides a good balance between latency to the data and good scalability; see section 3.2.3 for a deeper analysis in the difference between a centralized index and a distributed one.

Figure 1 shows the Falkon architecture, including both the data management and data-aware scheduler components. We start with a user which submits tasks to the Falkon wait queue. The wait queue length triggers the dynamic resource provisioning to allocate resources via GRAM4 from the available set of resources, which in turn allocates the resources and bootstraps the executors on the remote machines. The black dotted lines represent the scheduler sending the task to the compute nodes, along with the necessary information about where to find input data. The red thick solid lines represent the ability for each executor to get data from remote persistent storage. The blue thin solid lines represent the ability for each storage resource to obtain cached data from another peer executor. (The current implementation runs a GridFTP server [36] alongside each executor, which allows other executors to read data from its cache.)

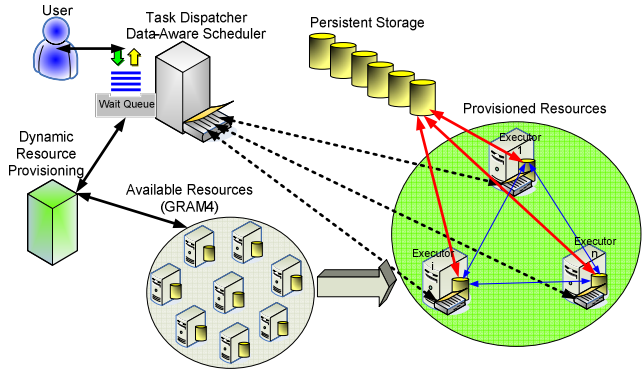

**Figure 1: Architecture overview of Falkon extended with data diffusion (data management and data-aware scheduler)**

### 3.2.2 Data Diffusion Execution Model

We assume that data is not modified after initial creation, an assumption that we found to be true for many data analysis applications. Thus, we can avoid complicated and expensive cache coherence schemes. We implement four well-known cache eviction policies [18]: *Random*, *FIFO* (First In First Out), *LRU* (Least Recently Used), and *LFU* (Least Frequently Used). The experiments in this paper all use LRU; we will study the effects of other policies in future work.

We also implement four task dispatch policies, as follows.

The **first-available** policy ignores data location information when selecting an executor for a task; it simply chooses the first available executor, and furthermore provides the executor with no information concerning the location of data objects needed by the task. Thus, the executor must fetch all data needed by a task from persistent storage on every access.

The **first-cache-available** policy selects executor for tasks in the same way as first-available; it differs in performing index lookups for each required data object and transferring the resulting location information along with the task description to the selected executor. Thus, the executor can fetch data needed by a task either from another executor, if cached there, or from persistent storage.

The **max-cache-hit** policy uses information about data location to dispatch each task to the executor with the largest number of data needed by that task. If that executor is busy, task dispatch is delayed until the executor becomes available. This strategy can be expected to reduce data movement operations compared to first-cache-available, but may lead to load imbalances, especially if data popularity is not uniform.

The **max-compute-util** policy also leverages data location information, but in a different way. It always sends a task to an available executor, but if there are several candidates, it chooses the one that has the most data needed by the task.

In each of the latter three cases, the centralized scheduler includes the necessary information to locate needed data (i.e., data stored in peer executor caches) without further lookups incurred at the executors. More details on the Falkon and data diffusion execution model are provided elsewhere [4, 30].

### 3.2.3 Centralized vs. Distributed Cache Index

Our central index and the separate per-executor indices are implemented as in-memory hash tables. The hash table implementation in Java 1.5 requires about 200 bytes per entry, allowing for index sizes of 8M entries with 1.5GB of heap, and 43M entries with 8GB of heap. Update and lookup performance on the hash table is good, with insert times in the 1~3 microseconds range (1M to 8M entries), and lookup times between 0.25 and 1 microsecond (1M to 8M entries). Thus, we can achieve an upper bound throughput of 4M lookups/sec.

In practice, the scheduler may make multiple updates and lookups per scheduling decision, and hence the effective scheduling throughput that can be achieved is lower. Falkon's non-data-aware scheduler (which simply does load balancing) can dispatch tasks at rates of 3800 tasks/sec on an 8-core system. In order for the data-aware scheduler to not become the bottleneck, it needs to make decisions within 2.1 milliseconds, which translates to over 3700 updates or over 8700 lookups to the hash table. Thus, we see that the rate-liming step remains the communication between the client, the service, and the executors.

Nevertheless, our centralized index could become saturated in a sufficiently large enough deployment. In that case, a more distributed index might perform and scale better. Such an index could be implemented using the peer-to-peer replica location service (P-RLS) [35] or distributed hash table (DHT) [31].

Chervenak et al. [35] report that P-RLS lookup latency for an index of 1M entries increases from 0.5 ms to just over 3 ms as the number of P-RLS nodes grows from 1 to 15 nodes. To compare their data with a central index, we present in Figure 2:

1) P-RLS performance data. Solid blue horizontal bars represent Chervenak et al.'s data; from 1 to 15 nodes; solid red horizontal bars represent predictions using a logarithmic best-fit curve, from 16 to 1M. nodes.

2) The predicted aggregate P-RLS throughput, in lookups/sec, based on the observed and predicted latency numbers, (The blue curve with red dots.)

3) The throughput achieved using the central index running on a single node, in lookup/sec. (The horizontal black line.)

We see that although P-RLS latencies do not increase significantly with number of nodes (from 0.5 ms with 1 node to 15 ms with 1M nodes), a considerable number of nodes are required to match that of an in-memory hash table. P-RLS would need more than 32K nodes to achieve an aggregate throughput similar to that of an in-memory hash table, which is 4.18M lookups/sec. In presenting these results we do not intend to argue that we need 4M+ lookups per second to maintain 4K scheduling decisions per second. However, these results do lead us to conclude that a centralized index can often perform better than a distributed index.

There are two disadvantages to our centralized index. The first is the requirement that the index fit in memory. Single SMP nodes can be bought with 128GB of memory, which would allow 683M entries in the index. However, this might not suffice for large applications. The second disadvantage is the single point of failure. Note that other elements of the Falkon service are also centralized, so distributing the index will only remove the single point of failure. We will investigate approaches to distributing the entire Falkon service to alleviate these two limitations.

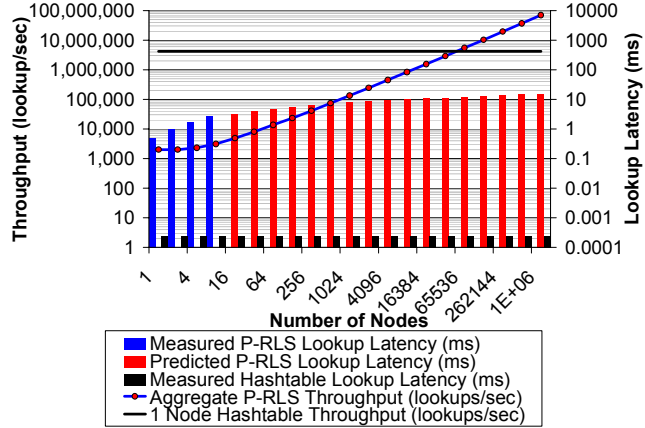

**Figure 2: P-RLS vs. Hash Table performance for 1M entries (P-RLS data come from Chervenak et al. [35])**

## 4. MICRO-BENCHMARKS

This section describes our performance evaluation of data diffusion using micro-benchmarks.

### 4.1 Testbed Description

Table 1 lists the platforms used in experiments. The UC_x64 node was used to run the Falkon service, while the TG_ANL_IA32 and TG_ANL_IA64 clusters [24] were used to run the executors. Both clusters are connected internally via Gigabit Ethernet, and have a shared file system (GPFS) mounted across both clusters that we use as the "persistent storage" in our experiments. The GPFS file system has 8 I/O nodes to handle the shared file system traffic. We assume a one-to-one mapping between executors and nodes in all experiments. Latency between UC_x64 and the compute clusters was between one and two ms.

**Table 1: Platform descriptions**

| Name | # of Nodes | Processors | Memory | Network |
|---|---|---|---|---|
| TG_ANL_IA32 | 98 | Dual Xeon 2.4 GHz | 4GB | 1Gb/s |
| TG_ANL_IA64 | 64 | Dual Itanium 1.3 GHz | 4GB | 1Gb/s |
| UC_x64 | 1 | Dual Xeon 3GHz w/ HT | 2GB | 100Mb/s |

### 4.2 File System Performance

To understand data diffusion costs, we first study GPFS performance in the ANL/UC TG cluster on which we conducted all experiments. We performed 160 different experiments involving, in aggregate, 19.8M files, the transfer of 3.7TB of data, and 163 CPU hours. Due to space limitations, we only summarize these results here; details are in a technical report [32].

GPFS read performance tops out at 3.4Gb/s for large files (1GB), and achieves 75% of peak bandwidth with files as small as 1MB if enough nodes access GPFS concurrently. The performance increase beyond 8 nodes is only apparent for small files (1B to 1MB); for large files, the difference is small (<6% improvement from 8 to 64 nodes). It appears that 8 compute nodes are enough to saturate the 8 GPFS I/O servers given large enough files. Read+write performance tops out at 1.1Gb/s, and there is little gain from having more than 8 nodes access GPFS concurrently, except for small files.

In contrast, aggregate local disk access speed scales linearly with the number of nodes involved, and thus can reach much higher rates when many nodes are used. Using all 162 nodes of the two TG_ANL clusters, read throughput reaches 76Gb/s and read+write throughput reaches 25Gb/s: both around 22 times faster than GPFS. This performance differential is a great motivator for applications to favor the use of local disks over shared disks, especially as applications scale beyond the capabilities of the statically configured I/O servers used to service the shared file systems.

### 4.3 Data Diffusion Performance

We measured performance for eight configurations, two variants (read and read+write), seven node counts (1, 2, 4, 8, 16, 32, 64), and eight file sizes (1B, 1KB, 10KB, 100KB, 1MB, 10MB, 100MB, 1GB), for a total of 896 experiments. For all experiments (with the exception of the 100% data locality experiments where the caches were warm), data was initially located only on persistent storage, which in our case was GPFS.

The eight configurations are:

1. **Model (local disk)**: local disk performance
2. **Model (persistent storage)**: GPFS performance
3. **Falkon (first-available)**: Falkon using first-available task dispatch policy (see Section 3.2.2).
4. **Falkon (first-available) + Wrapper**: the same as (3), except that all task executions are performed via a wrapper similar to that used in many applications to create a sandbox execution environment. The wrapper script creates a temporary scratch directory on persistent storage, makes a symbolic link to the input file(s), executes the task, and finally removes the temporary scratch directory from persistent storage, along with any symbolic links
5. **Falkon (first-cache-available; 0% locality)**: Falkon using first-cache-available task dispatch policy, and with a workload that does not repeat any files, and hence produces 0% cache hits. That is, all files are read from persistent storage to local disk, and then the operations are performed on the local data.
6. **Falkon (first-cache-available; 100% locality)**: the same as (5), but with the local disk caches first populated (not as part of the timed experiment), and then the workload from (5) repeated four times. Thus, we could achieve cache hit rates as high as 100% as the total requested data fits in available cache space.
7. **Falkon (max-compute-util; 0% locality)**: identical to (5), but using max-compute-util rather than first-cache-available policy.
8. **Falkon (max-compute-util; 100% locality)**: identical to (6), but using max-compute-util rather than first-cache-available policy.

Figure 3 shows read throughput for 100MB files, seven of the eight configurations, and varying numbers of nodes. Configuration (8) has the best performance: 61.7Gb/s with 64 nodes (~94% of ideal). Even the first-cache-available policy which dispatches tasks to executors without concern for data location performs better (~5.7Gb/s) than the shared file system alone (~3.1Gb/s) when there are more than 16 nodes. With eight or less nodes, data-unaware scheduling with 100% data locality performs worse than GPFS (note that GPFS also has eight I/O servers); one hypothesis is that data is not dispersed evenly among the caches, and load imbalances reduce aggregate throughput, but we need to investigate further to better understand the performance of data-unaware scheduling at small scales.

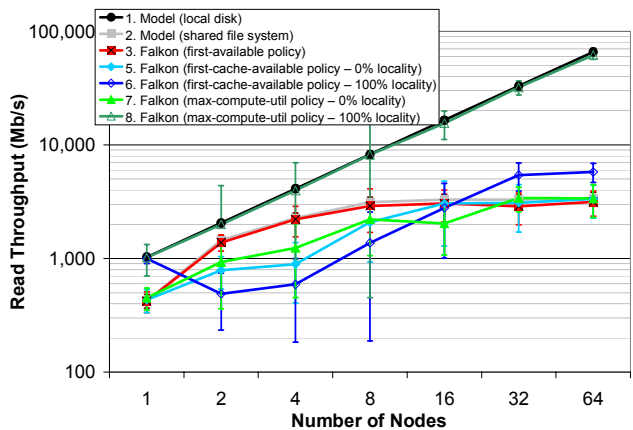

**Figure 3: Read throughput (Mb/s) for large files (100MB) for seven configurations for 1 – 64 nodes**

Figure 4 shows read+write performance, which is also good for the max-compute-util policy, yielding 22.7Gb/s (~96% of ideal). Without data-aware scheduling, throughput is 6.3Gb/s; when simply using persistent storage, it is a mere 1Gb/s.

In Figure 3 and 4, we omit configuration (4) as it had almost identical performance to configuration (3). Recall that configuration (4) introduced a wrapper script that created a temporary sandbox for the application to work in, and afterwards cleaned up by removing the sandbox. The performance of these two configurations was so similar here because of the large file sizes (100MB) used, which meant that the cost to create and remove the sand box was amortized over a large and expensive operation.

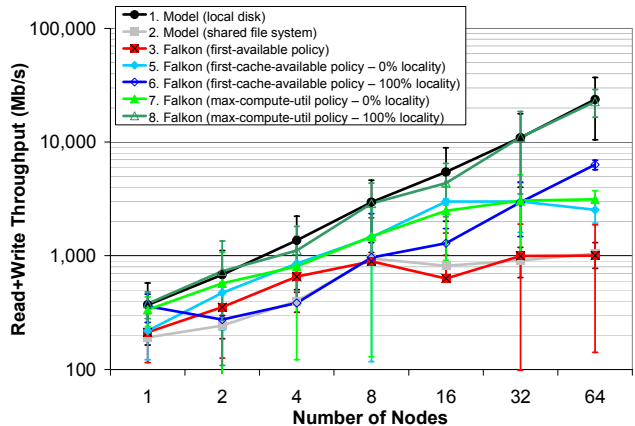

**Figure 4: Read+Write throughput (Mb/s) for large files (100MB) for seven configurations and 1 – 64 nodes**

Things look different when we consider smaller files. For example, Figure 5 shows read and read+write performance on 64 nodes for file sizes ranging from 1B to 1GB. Notice that for small file sizes (1B to 10MB), configuration (4) had one order of magnitude lower throughput than configurations (2) and (3). We find that the best throughput that can be achieved by 64 concurrent nodes with small files is 21 tasks/sec. The limiting factor is the need, for every task, to create a directory on persistent storage, create a symbolic link, and remove the directory. Many applications that use persistent storage to read and write files from many compute processors use this method of a wrapper to cleanly separate the data between different application invocations. This offers further example of how GPFS performance can significantly impact application performance.

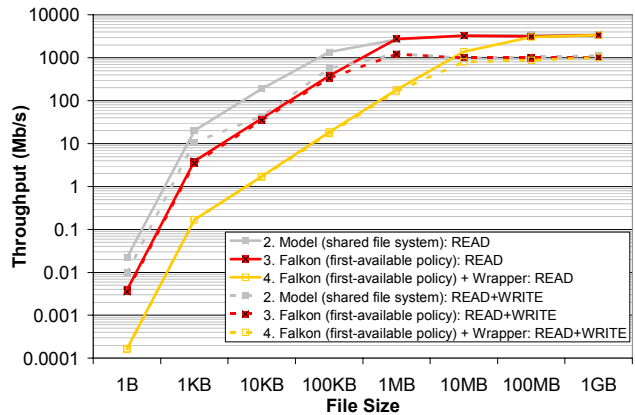

**Figure 5: Read and Read+Write throughput (Mb/s) for a wide range of file sizes for three configurations on 64 nodes**

Further results relating to Figure 3, 4, and 5 are in a technical report [32]. Overall, the shared file system seems to offer good performance for up to eight concurrent nodes (mostly due to there being eight I/O nodes servicing GPFS), however when more than eight nodes require access to data, the data diffusion mechanisms significantly outperform the persistent storage system. The improved performance can be attributed to the linear increase in I/O bandwidth with compute nodes, and the effective data-aware scheduling performed.

## 5. IMAGE STACKING IN ASTRONOMY

Prior to the work presented in this paper, we had assumed a shared file system was used for all data access. This approach works well for non-data intensive applications, but has scaling problems when dealing with large datasets and with particular data access patterns (many random small I/O reads/writes, and/or data intensive access patterns) on the shared file system. Our experience with astronomy specific data access patterns on TeraGrid [20] has been that performance of processing data directly from local disk as opposed to accessing the data from shared storage resources (i.e., GPFS [8]) can produce an order of magnitude difference [5, 6, 22].

We also evaluate the performance of our data diffusion mechanism in a real application. The application in question involves the "stacking" of image cutouts from different parts of the sky, with the goal of improving signal-to-noise for faint objects. Astronomical image collections usually cover an area of sky several times (in different wavebands, different times, etc). On the other hand, there are large differences in the sensitivities of different observations: objects detected in one band are often too faint to be seen in another survey. In such cases we still would like to see whether these objects can be detected, even in a statistical fashion. There has been a growing interest in re-projecting each image to a common set of pixel planes, then coadding many images to obtain a detectable signal that can to measure their average brightness/shape, etc. While this method has been applied for years manually for a small number of images, performing it on wide areas of sky in a systematic way has not yet been tried. It is also expected that much fainter sources (e.g., transient objects) can be detected from stacked images than can be detected in any individual image.

## 5.1 Workload Characterization

Astronomical surveys produce terabytes of data, and contain millions of objects. For example, the SDSS DR5 dataset (which we base our experiments on) has 320M objects in 9TB of images [9]. To construct realistic workloads, we identified the interesting objects (for a quasar search) from SDSS DR5; we used the CAS SkyServer [25] to issue the SQL command from Figure 6. This query retrieved 168,529 objects, which after removal of duplicates left 154,345 objects per band (there are 5 bands, u, g, r, I, and z) stored in 111,700 files per band.

```
select SpecRa, SpecDec
from QsoConcordanceAll
where bestMode=1
  and SpecSciencePrimary=1
  and SpecRa<>0
```

**Figure 6: SQL command to identify interesting objects for a quasar search from the SDSS DR5 dataset**

The entire working set consisted of 771,725 objects in 558,500 files, where each file was either 2MB compressed or 6MB uncompressed, resulting in a total of 1.1TB compressed and 3.35TB uncompressed. From this working set, various workloads were defined (see Table 2) that had certain data locality characteristics, varying from the lowest locality of 1 (i.e., 1-1 mapping between objects and files) to the highest locality of 30 (i.e., each file contained 30 objects on average of).

**Table 2: Workload characteristics**

| Locality | Number of Objects | Number of Files |
|---|---|---|
| 1 | 111700 | 111700 |
| 1.38 | 154345 | 111699 |
| 2 | 97999 | 49000 |
| 3 | 88857 | 29620 |
| 4 | 76575 | 19145 |
| 5 | 60590 | 12120 |
| 10 | 46480 | 4650 |
| 20 | 40460 | 2025 |
| 30 | 23695 | 790 |

## 5.2 Stacking Code Profiling

We first profile the stacking code to see where time is spent. We partition time into four categories, as follows.

1. *open:* open Fits file for reading
2. *radec2xy:* convert coordinates from RA DEC to X Y
3. *readHDU+getTile+curl+convertArray*:
    a. *readHDU:* reads header and image data
    b. *getTile:* perform extraction of ROI from memory
    c. *curl:* convert the 1-D pixel data (as read from the image file) into a 2-dimensional pixel array
    d. *convertArray:* convert the ROI from having SHORT value to having DOUBLE values
4. *calibration+interpolation+doStacking*:
    a. *calibration:* apply calibration on ROI using the SKY and CAL variables
    b. *interpolation:* do the appropriate pixel shifting to ensure the center of the object is a whole pixel
    c. *doStacking:* perform the stacking of ROI that are stored in memory
5. *writeStacking:* write the stacked image to a file

To simplify experiments, we perform tests with a simple standalone program on 1000 objects of 100x100 pixels, and repeat each measurement 10 times, each time on different objects residing in different files. In Figure 7, the Y-axis is time per task per code block measured in milliseconds (ms). Having the image data in compressed format affects the time to stack an image significantly, increasing the time needed by a factor of two. Similarly, accessing the image data from local disk instead of the shared file system speeds up processing 1.5 times. In all cases, the dominant operations are file metadata and I/O operations. For example, calibration, interpolation, and doStacking take less than 1 ms in all cases. Radec2xy consumes another 10~20% of total time, but the rest is spent opening the file and reading the image data to memory. In compressed format (GZ), there is only 2MB of data to read, while in uncompressed format (FIT) there are 6MB to read. However, uncompressing images is CPU intensive, and in the case of a single CPU, it is slower than if the image was uncompressed. In the case of many CPUs, the compressed format is faster mostly due to limitations imposed by the shared file system. Overall, Figure 7 shows the stacking analysis to be I/O bound and data intensive.

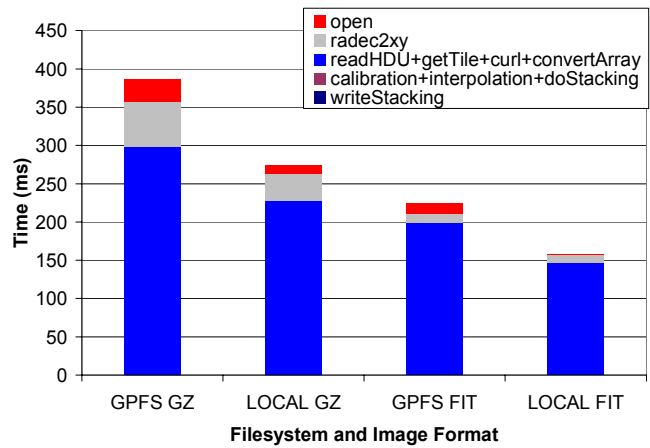

**Figure 7: Stacking code performance profiling for 1 CPU**

## 5.3 Performance Evaluation

All tests performed in this section were done using the testbed described in Table 1, using from 1 to 64, and the workloads (described in Table 2) that had locality ranging from 1 to 30. The experiments investigate the performance and scalability of the stacking code in four configurations: 1) Data Diffusion (GZ), 2) Data Diffusion (FIT), 3) GPFS (GZ), and 4) GPFS (FIT). At the start of each experiment, all data is present only on the persistent storage system (GPFS). In the data diffusion experiments, we use the max-compute-util policy and cache data on local nodes. For the GPFS experiments we use the next-available policy and perform no caching. GZ indicates that the image data is in compressed format while FIT indicates that the image data is uncompressed.

Figure 8 shows the performance difference between data diffusion and GPFS when data locality is small (1.38). We normalize the results here by showing the time per stacking operation (as described in Section 5.2 and Figure 7) per CPU used; with perfect scalability, the time per stack should remain constant as we increase the number of CPUs.

We see in Figure 8 that data diffusion and GPFS perform quite similarly when locality is low, with data diffusion slightly faster; data diffusion has a growing advantage as the number of CPUs increases. This similarity in performance is not surprising because most of the data must still be read from GPFS to populate the local disk caches. Note that in with small number of CPUs, it is more efficient to access uncompressed data; however, as the number of CPUs increases, compressed data becomes preferable. A close inspection of the I/O throughput achieved (not shown for space reasons) reveals that GPFS becomes saturated at around 16 CPUs with 3.4Gb/s read rates. In the compressed format (which reduces the amount of data that needs to be transferred from GPFS by a factor of three), GPFS only becomes saturated at 128 CPUs. We also find that when working in the compressed format, it is faster (as much 32% less per stack time) to first cache the compressed files, uncompress the files, and work on the files in uncompressed format, as opposed to working directly on the uncompressed files from GPFS.

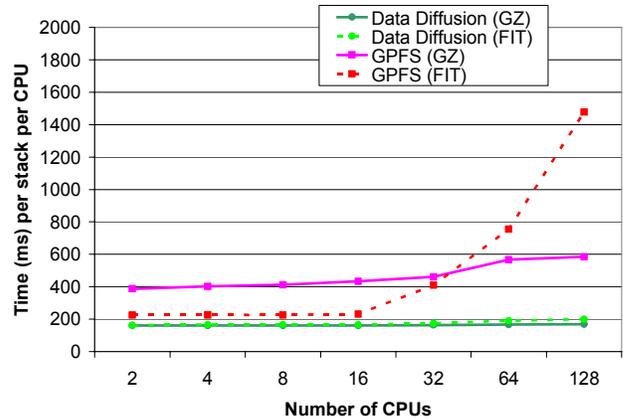

**Figure 9: Performance of the stacking application for a workload data locality of 30 using data diffusion and GPFS while varying the CPUs from 2 to 128**

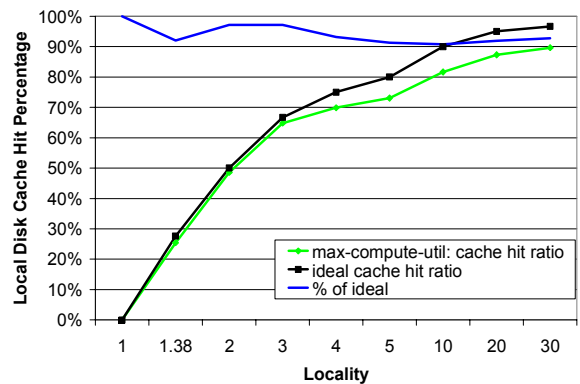

**Figure 10: Cache hit performance of the data-aware scheduler for the stacking application using 128 CPUs for workloads ranging from 1 to 30 data locality using data diffusion**

The following experiment (Figure 11) offers a detailed view of the performance (time per stack per CPU) of the stacking application as we vary the locality. The last data point in each case represents ideal performance when running on a single node. Note that although the GPFS results show improvements as locality increases, the results are far from ideal. However, we see data diffusion gets close to the ideal as locality increases beyond 10.

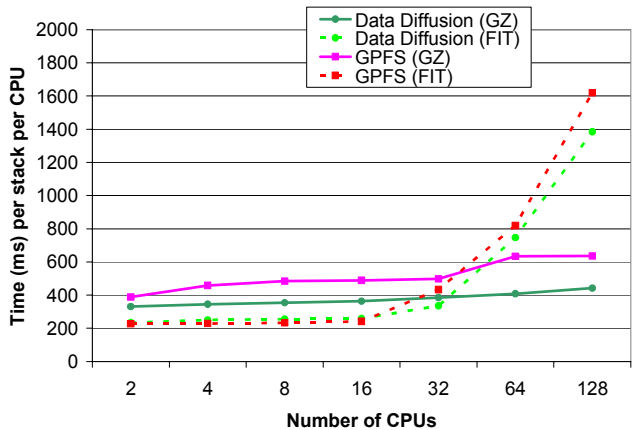

**Figure 8: Performance of the stacking application for a workload data locality of 1.38 using data diffusion and GPFS while varying the CPUs from 2 to 128**

While the previous results from Figure 8 shows an almost worst case scenario where the data locality is small (1.38), the next set of results (Figure 9) shows a best case scenario in which the locality is high (30). Here we see an almost ideal speedup (i.e., a flat line) with data diffusion in both compressed and uncompressed formats, while the GPFS results remain similar to those presented in Figure 8.

Data diffusion can make its largest impact on larger scale deployments, and hence we ran a series of experiments to capture the performance at a larger scale (128 CPUs) as we vary the data locality. We investigated the data-aware scheduler's ability to exploit the data locality found in the various workloads and its ability to direct tasks to computers on which needed data was cached. We found that the data-aware scheduler can get within 90% of the ideal cache hit ratios in all cases (see Figure 10). The ideal cache hit ratio is computed by $1 - 1/locality$; for example, with locality 3 (meaning that each file is access 3 times, one cache miss, and 2 cache hits), the ideal cache hit ratio is $1 - 1/3 = 2/3$.

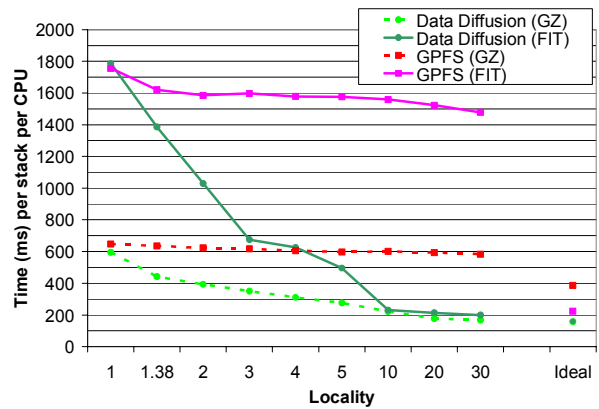

**Figure 11: Performance of the stacking application using 128 CPUs for workloads with data locality ranging from 1 to 30, using data diffusion and GPFS**

Figure 12 shows aggregate I/O throughput and data movement for the experiments of Figure 11. The two dotted lines show I/O throughput when performing stacking directly against GPFS: we achieve 4Gb/s with a data locality of 30. The data diffusion I/O throughput is separated into three distinct parts: 1) local, 2) cache-to-cache, and 3) GPFS, as a stacking may read directly from local disk if data is cached on the executor node, from a remote cache if data is on other nodes, and from GPFS as some data may not have been cached at all.

GPFS throughput is highest with low locality and lowest with high locality; the intuition is that with low locality, the majority of the data must be read from GPFS, but with high locality, the data can be mostly read locally. Note that cache-to-cache throughput increases with locality, but never grows significantly; we attribute this result to the good performance of the data-aware scheduler, always gets within 90% of the ideal cache hit ratio (for the workloads presented in this paper). Using data diffusion, we achieve an aggregated I/O throughput of 39Gb/s with high data locality, a significantly higher rate than with GPFS, which tops out at 4Gb/s.

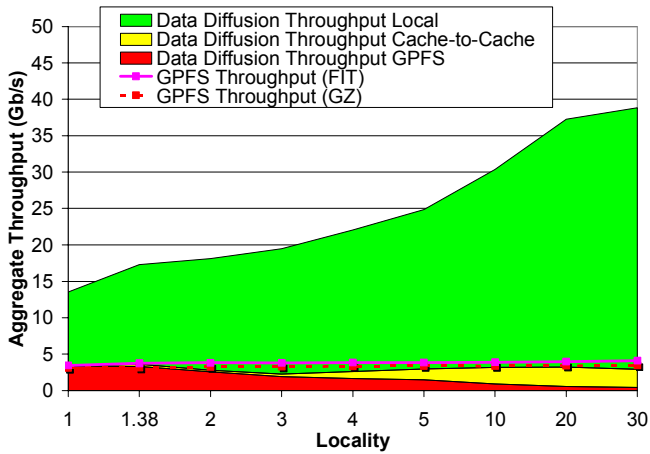

**Figure 12: I/O throughput of the stacking application using 128 CPUs, for workloads with data locality ranging from 1 to 30, and using both data diffusion and GPFS**

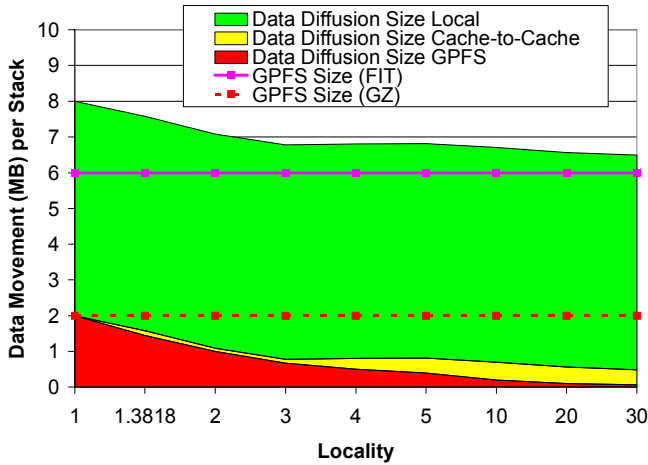

**Figure 13: Data movement for the stacking application using 128 CPUs, for workloads with data locality ranging from 1 to 30, using data diffusion and GPFS**

Finally, Figure 13 investigates the amount of data movement that occurs per stacking as we vary data locality. In summary, data diffusion (using compressed data) transfers a total of 8MB (2MB from GPFS and 6MB from local disk) for a data locality of 1; if data diffusion is not used, we need 2MB if in compressed format, or 6MB in uncompressed format, but this data must come from GPFS. As data locality increases, data movement from GPFS does not change (given a large number of CPUs and the small probability of data being re-used without data-aware scheduling). However, with data diffusion, the amount of data movement decreases substantially from GPFS (from 2MB with a locality of 1 to 0.066MB with a locality of 30), while cache-to-cache increases from 0 to 0.421MB per stacking respectively. These results show the decreased load on shared infrastructure (i.e., GPFS), which ultimately allows data diffusion to scale better.

## 6. CONCLUSIONS

Dynamic analysis of large datasets is becoming increasingly important in many domains. When building systems to perform such analyses, we face difficult tradeoffs. Do we dedicate computing and storage resources to analysis tasks, enabling rapid data access but wasting resources when analysis is not being performed? Or do we move data to computers when analysis requests occur, incurring expensive data transfer costs?

We describe here a *data diffusion* approach to this problem that seeks to combine elements of both dedicated and on-demand approaches. The key idea is that we respond to demands for data analysis by allocating data and compute systems and migrating code and data to those systems. We then retain these dynamically allocated resources (and cached code and data) for some time, so that if workloads feature data locality, they will obtain the performance benefits of dedicated resources.

To explore this approach, we have extended the Falkon dynamic resource provisioning and task dispatch system to cache data at executors and incorporate data-aware scheduling policies at the dispatcher. In this way, we leverage the performance advantages of high-speed local disk and reduce access to persistent storage.

Results from both micro-benchmarks and an astronomy image stacking application show that our approach can improve performance relative to alternative approaches. The performance benefits increase with the number of nodes used, as aggregate local I/O bandwidth scales linearly with the number of executors.

In future work, we plan to explore more sophisticated algorithms that address, for example, what happens when an executor is released; should we discard cached data, should it be moved to another executor, or should it be moved to persistent storage; do cache eviction policies affect cache hit ratio performance? Answers to these and other related questions will presumably depend on workload and system characteristics.

We also plan to use the Swift parallel programming system to explore data diffusion performance with more applications and workloads. We have integrated Falkon into the Karajan workflow engine used by Swift [16, 33]. Thus, Karajan and Swift applications can use Falkon without modification. Swift has been applied to applications in the physical sciences, biological sciences, social sciences, humanities, computer science, and science education. We have already run several applications (fMRI, Montage, MolDyn) without data diffusion [4, 16, 33], on which we will investigate the benefits of data diffusion as well.

## 7. ACKNOWLEDGEMENTS
This work was supported in part by the NASA Ames Research Center GSRP Grant Number NNA06CB89H and by the Mathematical, Information, and Computational Sciences Division subprogram of the Office of Advanced Scientific Computing Research, Office of Science, U.S. Dept. of Energy, under Contract DE-AC02-06CH11357. We also thank TeraGrid and the Computation Institute at University of Chicago for hosting the experiments reported in this paper.